\documentstyle[prl,twocolumn,aps,epsf]{revtex}
\begin{document}

\noindent {\bf
Comment on ``One-Dimensional Disordered \\
 Bosonic Hubbard Model: A Density-Matrix \\
 Renormalization Group Study"}

\bigskip
In a recent letter Pai {\sl et al}. \cite{Pai} have applied the
density-matrix renormalization group (DMRG) method to the study of
quantum phase transitions in interacting disordered Bose system.
They claim to obtain an accurate phase diagram for the
Hamiltonian
\begin{equation}
{{\cal H} \over t} = -\sum_{<ij>} a_i^{\dag}a_j +
U\sum_{i} n_i(n_i-1)- 2 \sum_{i} \mu_i n_i \;,
\end{equation}
containing hopping term, on-site repulsion and
random potential $\mu_i$ uniformly distributed between
$-\Delta$ and $\Delta$.
The authors proved that the method works remarkably well in pure
system and allows precise calculation of the phase
transition point between the superfluid (SF) and Mott-insulator (MI)
phases. This was considered as a sufficient ground to apply DMRG method
to the disordered system.
In this comment we agrue that DMRG completely fails in predicting
 phase diagram in disordered case.
At considerable disorder it has wrong
shape  and is completely incorrect quantitatively, as proved by
quantum Monte Carlo simulations.  For weak disorder
the phase diagram obtained in Ref.~\cite{Pai} contradicts
 theoretical predictions (and an exact theorem!).

\bigskip
\begin{figure}
\epsfxsize=6.6cm
\epsffile{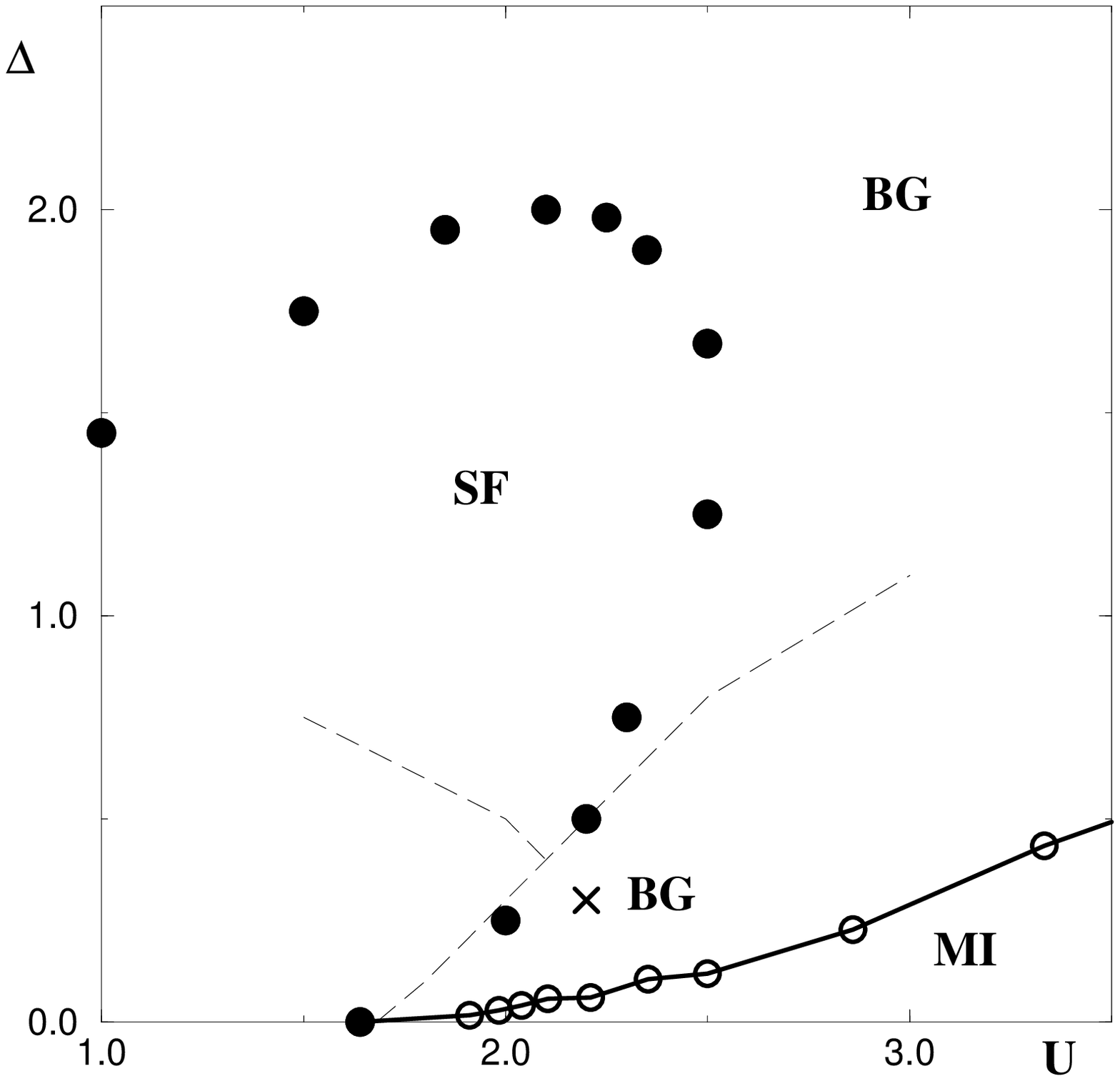}
\vspace{-3.4cm}
{\caption ~
The phase diagram for the Hamiltonian (1). The error bars
are within the points size and originate mostly from the finite-size
corrections (the system size used in this calculations was $L=100$
as in Ref.~\cite{Pai}).
BG phase at the point maked by the cross was verified for the  system
with $L=1000$ and $t/T =200$.
Dashed line indicates the results obtained by the DMRG method \cite{Pai}.}
\label{f1}
\end{figure}

Recent developments in the quantum Monte Carlo (QMC) algorithms
allow one to calculate directly the superfluid stiffness, compressibility
and the dependence of the number of particles on the chemical potential
in systems as large as few hundred lattice sites \cite{our}.
In Fig. 1  we present the phase diagram for the Hamiltonian (1)
obtained by the QMC simulations.
The phase boundary between the SF and Bose glass (BG)
regions was found from the condition $K^{-1}=\pi( \kappa \rho_s)^{1/2} =3/2$,
where $\kappa$ is compressibility, $\rho_s$ is the superfluid stiffness.
The critical value for the exponent $K$ at the SF-BG transition is
known \cite{Giamarchi} to be $2/3$. We see that the actual phase diagram
has little in common with that found in Ref.~\cite{Pai}, except at weak
disorder where the phase boundary SF-BG was incorrectly identified with
the SF-MI transition. To this end we note that simple considerations
\cite{Freericks}
predict the upper bound for the MI-BG transition at $\Delta = E_{gap}/4$
where $E_{gap}$ is the energy gap in a pure system. The corresponding
line is shown in Fig.1 by open circles  utilizing the data for $E_{gap}$ found
in Ref.~\cite{Kash}. As expected, it is well below  the SF-BG line \cite{Svist}.
We have explicitly
checked that the state above the solid line at $U=2.2$
is a gapless state with almost zero superfluid stiffness.

In our opinion, the problem with the DMRG method in disordered case
arises because it implies that density matrix of
any sufficiently large system block of $R$ sites ($R>>1$) is
``representative", i.e., describes a typical system behavior
on the scale $R$; the variation of boundary conditions is supposed to be small
when slightly increasing the block size.
This assumption flatly contradicts the physics of strongly disordered case,
 the MI-BG transition near the critical point, and  the  SF-BG transition
in weakly interacting system.
In the last two cases the large-scale behavior
is governed by exponentially rare Lifshitz regions
of strong disorder, where the system wave-function changes drastically
on rather small length scale, while at strong disorder
such variations of the wave function are just typical.
DMRG method, working with finite number
of basis states fails to accomodate the sudden change
in the boundary conditions when the block size is slightly increased,
simply because the number of states kept is too small to reconstruct
the correct new basis set.

\bigskip
\noindent N. V. Prokof'ev and B. V. Svistunov \\
\indent Russian Research Center ``Kurchatov Institute",\\
\indent Moscow 123182, Russia


\begin{thebibliography}{99}

\vspace{-1.5cm}
\bibitem{Pai} R.V. Pai, R.P. Pandit, H.R. Krishnamurthy, and
S. Ramasesha, Phys. Rev. Lett., {\bf 76}, 2937 (1996).

\bibitem{our} 
      N.V. Prokof'ev, B.V. Svistunov, and I.S. Tupitsyn, Pis'ma v Zh. Eksp.
      Teor. Fiz. {\bf 64}, 853 (1996); cond-mat/9703200 (submitted
      to Phys. Rev. B).

\bibitem{Giamarchi} T. Giamarchi and H.J. Schultz, Europhys. Lett.,
{\bf 3}, 1287; Phys. Rev. B, {\bf 37}, 325 (1988).


\bibitem{Freericks}
   M.P.A. Fisher, P.B. Weichman, G. Grinstein, and D.S. Fisher,
   Phys. Rev. B, {\bf 40}, 546 (1989);
   J.K. Freericks and H. Monien,  Phys. Rev. B
   {\bf 53}, 2691 (1996).

\bibitem{Kash}
   V.A. Kashurnikov, A.V. Krasavin, and B.V. Svistunov, 
   Pis'ma v Zh. Eksp. Teor. Fiz. {\bf 64}, 92 (1996). 

\bibitem{Svist}
   B.V. Svistunov, 
   Phys. Rev. B, {\bf 54}, 13982 (1996). 

   

\end{thebibliography}
\end{document}